\documentclass[a4paper,twoside]{article}

\usepackage{epsfig}
\usepackage{subcaption}
\usepackage{calc}
\usepackage{amssymb}
\usepackage{amstext}
\usepackage{amsmath}
\usepackage{amsthm}
\usepackage{multicol}
\usepackage{pslatex}
\usepackage{xcolor}
\usepackage[bottom]{footmisc}
\usepackage{float}
\usepackage{hyperref} 
\hypersetup{ 
    colorlinks,
    linkcolor={red!30!black},
    citecolor={red!30!black},
    urlcolor={red!50!black}
}
\usepackage{SCITEPRESS}     
\usepackage{apalike}

\newcommand{\iMove}{MoveReminderApp}  
\newcommand{\REFramework}{RE4HCAI}  

\newcommand{\sectopic}[1]{\vspace{0.2em}\par\noindent{\textit{\bfseries #1}}}

\begin{document}

\title{Requirements Elicitation and Modelling of Artificial Intelligence Systems: An Empirical Study}

\author{\authorname{Khlood Ahmad\sup{1}, Mohamed Abdelrazek\sup{1}, Chetan Arora\sup{2}, John Grundy\sup{2} and Muneera Bano\sup{3}}
\affiliation{\sup{1}Deakin University, Geelong, VIC, Australia}
\affiliation{\sup{2}Monash University, Clayton,  VIC, Australia}
\affiliation{\sup{3}CSIRO's Data61, Clayton,  VIC, Australia}
\email{\{ ahmadkhl, mohamed.abdelrazek\}@deakin.edu.au, \{chetan.arora, john.grundy\}@monash.edu, muneera.bano@csiro.au}
}
\vspace{-3em}

\abstract{
Artificial Intelligence (AI) systems have gained significant traction in the recent past, creating new challenges in requirements engineering (RE) when building AI software systems. RE for AI practices have not been studied much and have scarce empirical studies. Additionally, many AI software solutions tend to focus on the technical aspects and ignore human-centered values.
In this paper, we report on a case study for eliciting and modeling requirements using our framework and a supporting tool for human-centred RE for AI systems. Our case study is a mobile health application for encouraging type-2 diabetic people to reduce their sedentary behavior. We conducted our study with three experts from the app team -- a software engineer, a project manager and a data scientist. We found in our study that most human-centered aspects were not originally considered when developing the first version of the application. We also report on other insights and challenges faced in RE for the health application, e.g., frequently changing requirements.}

\onecolumn \maketitle \normalsize \setcounter{footnote}{0} \vfill

\section{\uppercase{Introduction}}
\label{sec:introduction}

\vspace{-1em}
There has been a massive shift towards using Artificial Intelligence (AI) components in software solutions. This shift has been possible because of the increase in processing power, better AI models, and the availability of large datasets to train accurate AI models~\cite{holmquist2017intelligence}. However, incorporating AI components has impacted how we build software systems, and new issues have emerged in the software development process.

Existing methods and tools used in Software Engineering (SE) are often inadequate in building AI software~\cite{sculley2015hidden,feldt2018ways}. Most AI models rely heavily on the training data, have inherent inaccuracies, unpredictability, and are largely black-box in nature. These attributes have led to new requirements, usually not considered or emphasised in traditional SE, such as AI-based software ethics and data requirements~\cite{martinez2022software}. In traditional Requirements Engineering (RE), i.e. without an AI component, the process usually involves specifying requirements for systems that are deterministic, and the outputs are known early on. 
In RE for AI (RE4AI), it is more difficult to specify requirements for non-deterministic and black-box components, for which the outputs are usually unknown until the models are trained and tested with data~\cite{belani2019requirements,agarwal2014expert,khomh2018software}. \textcolor{black}{ However, requirements such as policies, business values and system constraints need to be established at early stages~\cite{ezzini23ai}.} 



AI-based software development has mostly focused on the technical aspects of the AI components~\cite{maguire2001methods,schmidt2020interactive} and overlooked human-centred aspects, such as age, gender, culture, emotions, ethnicity, and many others \cite{grundy2021impact,shneiderman2022human,Fazzini:22}. Lu et al.~\cite{lu2022software} interviewed 21 AI scientists and engineers and found that responsible AI requirements were overlooked or only included at very high-level system objectives. Overlooking these human-centred aspects when building AI software can lead to systems that are biased, discriminate against user groups, and are non-inclusive~\cite{amershi2014power}. 

In this paper, we build on our existing work on a structured literature review (SLR) on RE for AI systems~\cite{ahmad2021SLR,ahmad2022requirements}, a survey on practitioners' perspective on RE for AI systems~\cite{ahmad2023requirements},
an RE framework for eliciting and modelling human-centric AI software requirements. In this paper, we report on a case study we performed on a tool implementation of our framework on a mobile health application (\iMove) that encourages people with type-2 diabetes to move more. We shared the tool with three experts from the team working on the mobile application to elicit and model the requirements. We then interviewed the experts for feedback on the tool and framework and report on the findings in this paper. 

Our study found that most of the human-centred aspects were not considered in the development process. In RE for AI systems, it is difficult to specify certain requirement types at the early stages of the development process, as data requirements and model requirements usually have to be adjusted or changed for the model to achieve desirable outcomes. However, our framework can help address some of the issues during the early stages of the development process, e.g., addressing human-centred issues and thinking about the data and model requirements. Based on our findings from this case study, we propose that a progressive and iterative agile method could be more appropriate to adopt.  Furthermore, using our framework and the tool can help capture requirements changes as stakeholders learn more about the problem, data and potential models in the early stages of development. We note that RE for AI is a relatively new area of research, and empirical studies in this area are extremely scarce. This paper and our case study findings contribute to the empirical evidence on performing RE for AI.

The rest of the paper is structured as following:  Section~\ref{sec:related}
presented the related work. Section~\ref{sec:tool} presents details of our framework and the tool. Section~\ref{sec:case} reports on the details of our case study. Section~\ref{sec:elicitation} presents the outcomes of our sessions with experts on \iMove~study. Section~\ref{sec:discussion} discusses key results from the interviews and summarizes reflections from our study.  Section~\ref{sec:conclusion} concludes.

\vspace{-1em}

\section{\uppercase{Related Work}}~\label{sec:related}
\vspace{-2em}

\textbf{Requirements elicitation} is considered the most crucial part of RE as it sets to unravel and capture the need for the system from the stakeholders early on in the software development life cycle~\cite{davey2015requirements}. The process of eliciting requirements includes ``a set of activities that must allow for communication, prioritization, negotiation, and collaboration with all the relevant stakeholders''~\cite{zowghi2005requirements}. In order to capture the correct requirements, system boundaries need to be set and defined~\cite{nuseibeh2000requirements}.

The techniques used in eliciting requirements can vary from formal and informal interviews, surveys and questionnaires to other tasks such as scenarios, observations and protocol analysis~\cite{carrizo2014systematizing}. Several presenting issues related to requirements elicitation techniques include miscommunication and difficulties in transferring knowledge between the elicitor and stakeholder~\cite{fuentes2010understanding}. These issues involve eliciting: (1)~\textit{known unknowns} referring to knowledge of which the requirements engineer is aware of but not the stakeholder. (2)~\textit{unknown knowns} refers to the knowledge held by the stakeholder but not expressed to the requirements engineer. (3)~\textit{unknown unknowns} are the most challenging of all, wherein both the stakeholder and the requirements engineer are not aware of the existing knowledge in this situation~\cite{sutcliffe2013requirements,gervasi2013unpacking}. There are many unknown-unknowns in RE4AI, especially during the early stages of exploring AI software. Most AI-based projects lack a common understanding of what requirements are needed to build the project, with limited shared knowledge among the stakeholders and the team building the AI software, making it difficult to elicit and specify requirements. The unknowns include the outcomes of the system that may only be known once the model is trained on a given dataset. Thus, some requirements can only become known towards later stages of the software life-cycle.  

\textbf{Requirements modelling languages} are used to visually display requirements and identify the stakeholders' needs at a higher-level system abstraction~\cite{gonccalves2019understanding}. The examples of RE modelling languages include Goal-Oriented Requirements Languages (GRL)~\cite{lapouchnian2005goal} and the i* model~\cite{dalpiaz2016istar}. In GRL, goals are used to model non-functional requirements (NFR) and business rules~\cite{amyot2011user}. However, the disadvantage of using GLR is that it is difficult to learn among non-software engineers. Modelling languages such as UML and SysML have been used to model requirements in RE4AI. Although these modelling languages are easier to use and learn than GRL, they have limitations in modelling NFR's and business rules~\cite{ries2021mde,gruber2017integrated,amaral2020ontology}. Other studies proposed to use of conceptual models to present requirements in RE4AI~\cite{nalchigar2021modeling}. 

\textbf{Existing tools} used to support software engineers in developing human-centred AI software include the ``Human-AI INtegration Testing''(HINT) tool~\cite{chen2022hint} and the ``AI Playbook''~\cite{hong2021planning} that provides early prototyping to reduce potential errors and failures in AI software. 
Although these tools support building human-centred AI software, they are limited to the design phase and not to RE. Additionally, the tools found in the SLR, such as the GR4ML~\cite{nalchigar2021modeling} to visually present and model requirements for AI software systems did not focus on the human-centred aspects in their prototypes. 

\vspace{-1em}

\begin{figure*}[!t]
   \centering
\includegraphics[width=0.8\linewidth]{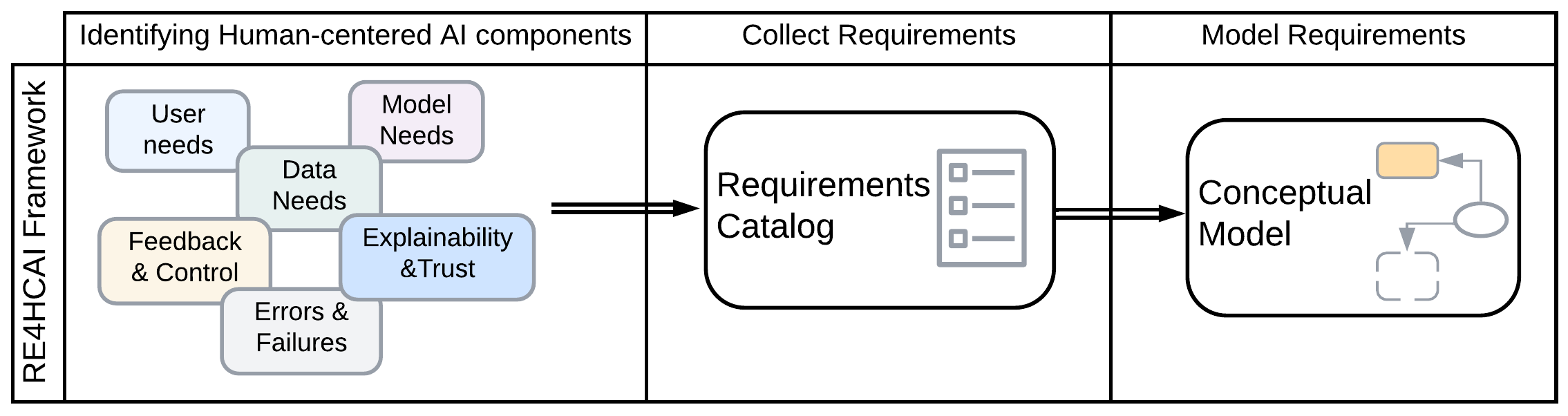}
\caption{Framework for eliciting and modeling requirements for human-centered AI software}
\label{fig:Framework}
\end{figure*}

\section{\uppercase{\REFramework~Framework}}~\label{sec:tool}
\vspace{-2em}

We proposed framework (\REFramework) for guiding the requirements engineers and other stakeholders in requirements elicitation and modelling of human-centered AI-based software. The framework has three layers, as shown in Figure~\ref{fig:Framework} and discussed below. 

\begin{figure*}[!t]
   \centering
\includegraphics[width=0.8\linewidth]{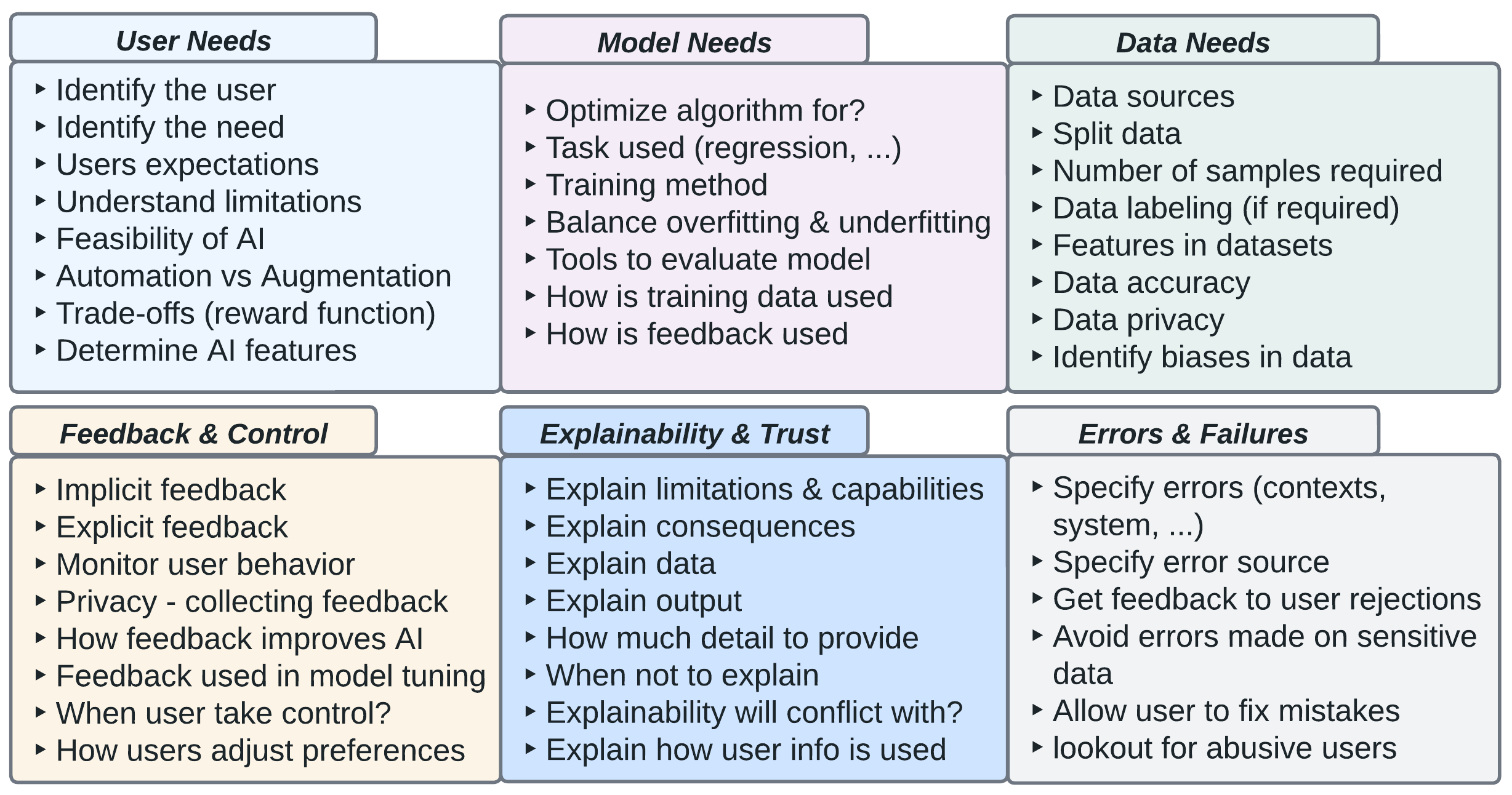}
\caption{Requirements for Human-Centred AI.}
\label{fig:HC-Needs}
\vspace*{-.6em}
\end{figure*}

\sectopic{Identifying Human-centered AI components.} The first layer consists of six categories of requirements that need to be elicited for developing human-centered AI software solutions. The six areas are User Needs, Model Needs, Data Needs, Feedback and User Control, Explainability and Trust, and Errors and Failure. The categories are inspired by Google PAIR's human-centred AI guidelines~\cite{GooglePair2019}. Our categorization into six areas aligns with the five categories proposed by Google PAIR (all areas except Model Needs). We added the `Model Needs' that identify the human-centered approaches used when selecting and training an appropriate AI model. Figure~\ref{fig:HC-Needs} shows a high-level summary of the requirements covered in each area.

\sectopic{Requirements Catalog.} For each of the six areas, we combined the human-centred guidelines from Google PAIR, Microsoft's guidelines for human-centred AI interaction~\cite{Microsoft2022}, and the guidelines for Apple's human interface for developing ML applications~\cite{Apple2020} along with the Machine Learning Canvas~\cite{MLCanvas}. Next, we then collected all requirements in the literature ~\cite{ahmad2022requirements,villamizar2021requirements} that fit our six selected areas and mapped them against the combined guidelines. To validate our mapping, we conducted a survey with practitioners to understand which of these six areas (and the guidelines in Figure~\ref{fig:HC-Needs} should be included in a~\REFramework~framework~\cite{ahmad2023requirements}. The results showed that all the six areas in the~\REFramework~framework are deemed important by practitioners when collecting AI requirements. The gathered human-centred requirements were listed in a tabular format to form our catalog, with six sections. Each section is dedicated to eliciting detailed requirements for each area of Figure~\ref{fig:HC-Needs}. 


\sectopic{Conceptual Model.} Our SLR reviewed papers that used a modeling language or notation to present requirements. We found that most of the studies preferred UML or a conceptual model. UML was a preferred choice because it is easier to understand and use among different groups. However, UML has some limitations when presenting requirements for AI and lacks direct support for modeling business rules and NFR's~\cite{ahmad2021SLR}. Also, in our~\REFramework~we have concepts that are difficult to visualize in UML, such as needs, limitations, and trade-offs. Therefore, we created a conceptual model which can be instantiated in a project to present the requirements visually.  

\begin{figure}[h]
   \centering
\includegraphics[width=\linewidth]{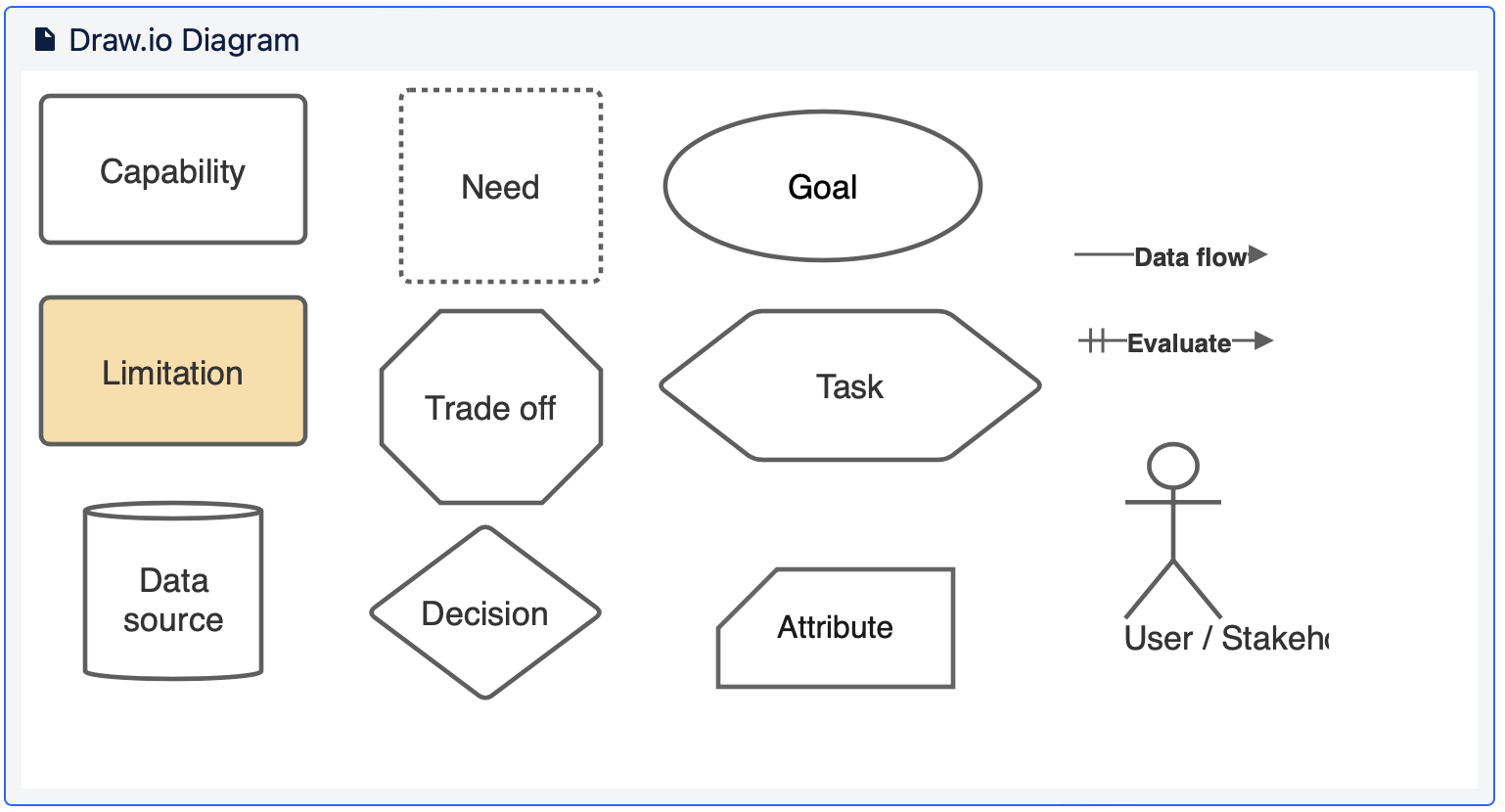}
\caption{Legend for the conceptual model at the second level to show more abstract requirements}
\label{fig:legend}
\end{figure}

Our modeling language consists of two layers. The first layer provides a holistic view of each of our six areas.
We show part of this model in Figure~\ref{fig:Level1-iMove}. In the first layer, we use UML class notations to showcase a high-level presentation of the requirements needed for each area and an oval shape to show the system's high-level goal and connect all six presented areas. The second layer presents a separate model for each area, and we use unified notations for all the areas as illustrated in Figure~\ref{fig:legend}.   When creating our modeling notations, we tried to adhere as much as possible to the Physics of Notations~\cite{moody2009physics}. We incorporated different shapes, colors, and textures to reduce cognitive overload for users and make the notations easy to use and understand. 
Each notation is used to model a different concept from the requirements collected in the requirements catalog.

\begin{figure*}[h]
   \centering
\includegraphics[width=\linewidth]{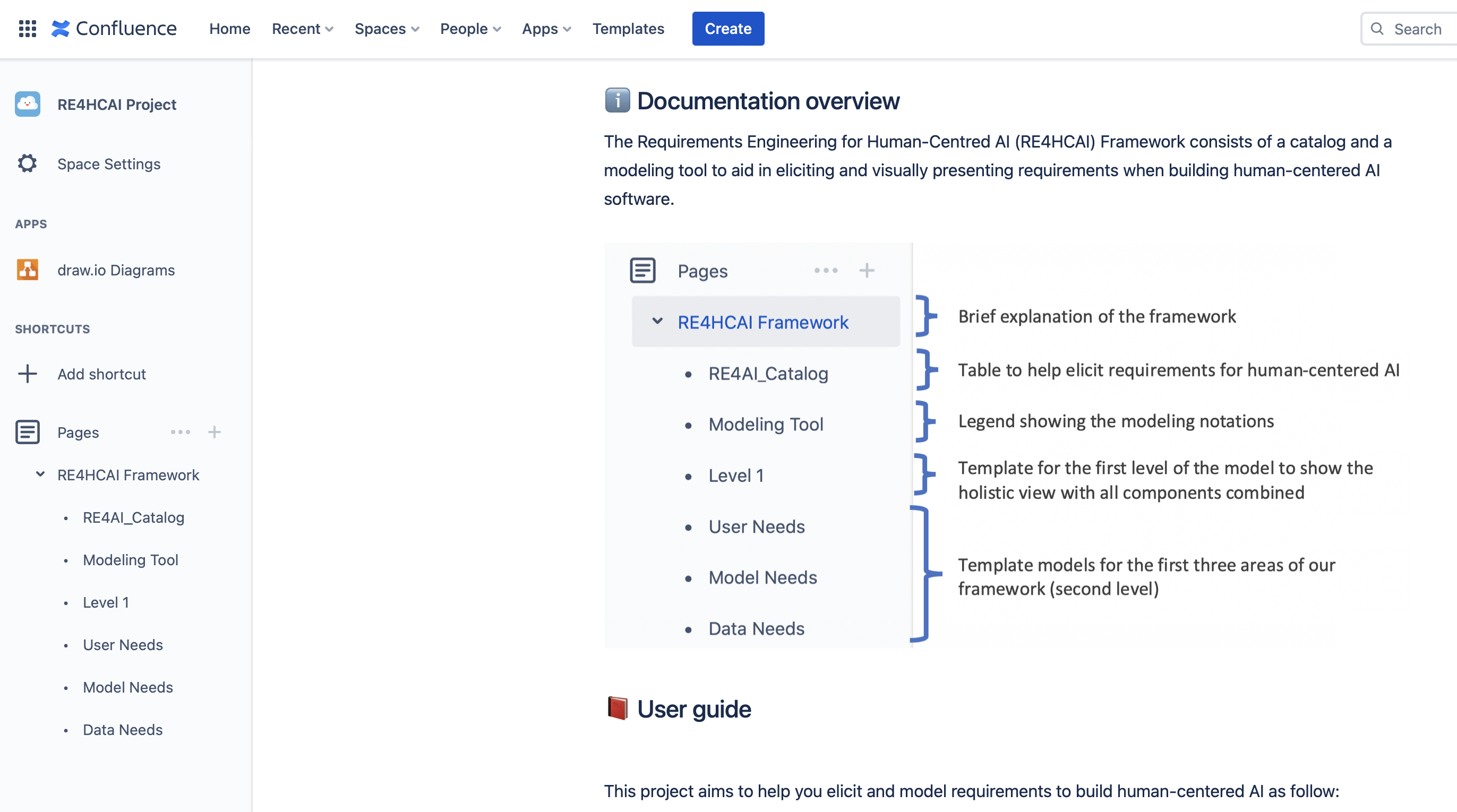}
\caption{The Confluence collaboration page used to conduct the case study}
\label{fig:Platform}
\end{figure*}

\subsection{\REFramework~Collaborative~Tool}
\vspace{-1em}
Developing a software system with an AI component requires different teams and roles to work together. In order to facilitate collaboration between the development teams, we aimed to develop a collaborative tool for implementing our framework. The idea was to use platforms that are familiar among different AI-software development roles, and utilize existing tools and collaboration platforms. As part of our practitioner survey~\cite{ahmad2023requirements}, we asked the participants about their platform preferences. Based on the survey results, we identified platforms to use for evaluating our initial prototype. We use Confluence ({\url{https://www.atlassian.com/software/confluence}), an online collaboration platform where team members can work on projects together and share, plan and build ideas together. Confluence is used as a mean to present our framework among participants and allows us to present our modelling tool using the existing drawing applications Draw.io ({\url{https://drawio-app.com}}).

The platform contains a template project space as shown in  Figure~\ref{fig:Platform}. The landing page provides an overview of the platform, framework, and a short tutorial on how to use the catalog and the modeling language. The main page further contains six sub-pages each dedicated to part of the framework. Sub-page one includes a table with the catalog for eliciting the human-centered requirements for a specific AI-software system. Sub-page two showcases the available modelling notations that are displayed in Figure~\ref{fig:legend}. Sub-page three contains the template model for the first layer model. The last three sub-pages provide templates for the models second layers. We only provide models for the first three areas at this stage, i.e., ``User Needs'', ``Model Needs'', and ``Data Needs'', and plan to expand it further in the future.
\vspace{-1.5em}

\section{\uppercase{Case study Details}}~\label{sec:case}
\vspace{-0.7cm}

We conducted a case study to investigate the usefulness of our framework and the corresponding platform to engineering AI software with a human-centered perspective. Our case study the \iMove~project \cite{daryabeygi2022development}, involved designing and building a real-time health application for people with type 2 diabetes (T2D). The application keeps track of the users' behavior toward movement, measures the sedentary time, and sends notifications to remind the user to move. These notifications consist of messages using a Deep Learning (DL) model to remind the person to stand up or walk once the user has been predicted as sitting for a prolonged time.

\begin{figure*}[h!]
   \centering
\includegraphics[width=\linewidth]{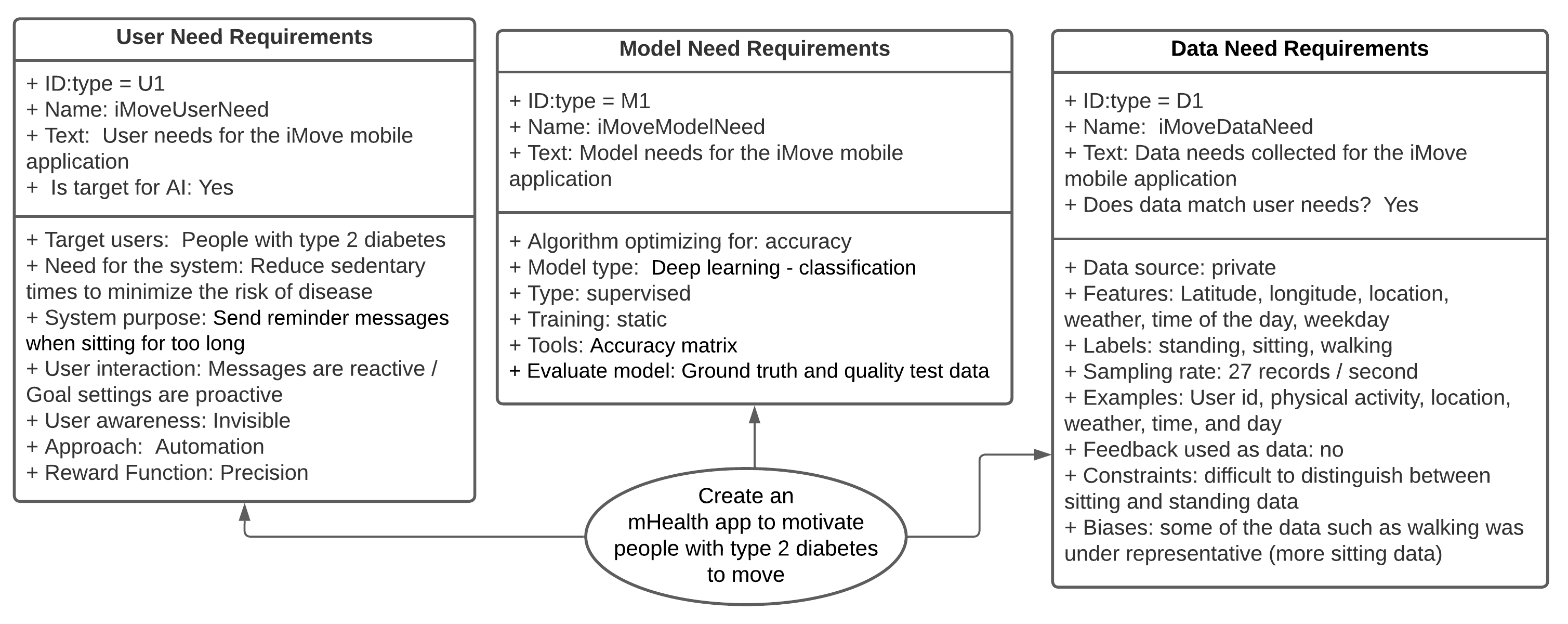}
\caption{Model presenting a holistic view of the requirements elicited for the \iMove~project.}
\label{fig:Level1-iMove}
\end{figure*}


We conducted our case study with three experts who worked on the \iMove~project, including a project manager, data scientist, and software engineer. Our participants have several years of professional experience: the project manager has nine years of experience, the software engineer has two years of experience, and the data scientist has four years of experience in their field. We note that getting these three experts in our case study provides complimentary views and perspectives from different roles in AI-based software development. Also, the \iMove~project was already in the late development stages and was intentionally selected as we wanted to investigate if the experts had initially considered different aspects of our framework while eliciting and specifying or modeling requirements.

\sectopic{Requirements Elicitation Sessions.} We created a dedicated project in Confluence for the \iMove~project, giving each participant access to the project. We first conducted a session with each participant. Due to the workplace recommendations and experts' preferences, all sessions were conducted virtually. The sessions were for 1-2 hours each. Each session was dedicated at making the participants aware of our framework and the Confluence platform and using their domain knowledge and expertise in the \iMove~project to investigate if our platform was valuable for their project's context. We further wanted to identify any discrepancies between the participants' views of the requirements needed for the first three sub-areas in our framework. To do so, we elicited requirements using the catalog from each of the three participants individually. 

Once all the requirements were elicited, a holistic view of the \iMove~project was established, as shown in Figure~\ref{fig:Level1-iMove}, and a model was created for each of the three areas as described in the next sections. We note that we used our requirements catalog of Figure~\ref{fig:HC-Needs} to elicit information for each of these three areas. Participants could return to the Confluence project at any time and edit the elicited requirements and models. 

\sectopic{Participant Interviews.} The final step in our study involved interviews with the participants to get their feedback on our platform and the framework. We note that the data scientist in the project was not available for the interview stage due to personal reasons, and only the software engineer and the project manager participated in our interviews. We obtained consent from both participant's, and interviews were carried out online. Due to the limited availability of the experts, the two interviews were scheduled for 45 minutes each. Both interviews were audio-recorded and transcribed for analysis and evaluation.

\vspace{-1em}

\section{\uppercase{Elicitation Sessions}}~\label{sec:elicitation}
\vspace{-0.05cm}
In this section, we provide details of the outcomes of our requirements elicitation and modeling sessions with the three participants. Below, we discuss the outcomes of eliciting requirements for the user needs, model needs, and data needs collated from the three sessions. Figure~\ref{fig:Level1-iMove} presents the high-level requirements for the first three areas and the main goal for the project.  When building the models, we used a blue tick to distinguish between framework requirements and system requirements, as shown in the first three models in Figures~\ref{fig:UserNeeds-iMove},~\ref{fig:DataNeeds-iMove}, and~\ref{fig:ModelNeeds-iMove}.  The notations that do not have a blue tick are specific to the \iMove~project.

\vspace{-0.2cm}
\subsection{\iMove: User Needs}~\label{sec:iMoveUserNeeds}
\vspace{-0.7cm}

We identify the need for the system as shown in part \textbf{a} (the blue colored box) in Figure~\ref{fig:UserNeeds-iMove}. The experts had considered, among others, previous scientific literature on the behavior of T2D patients to determine \emph{the need for the project}. A recent SLR showed that people with T2D have higher sedentary behavior, which can cause serious health risks~\cite{kennerly2018physical}. The existing literature found that getting people with T2D to break up their sitting time with little movements could improve quality of life and reduce increased health risks due to diabetes~\cite{dempsey2016sitting}. Using digital health solutions, such as sensors, wearables, and mobile phones, was found to reduce sitting times to 41 minutes per day~\cite{stephenson2017using}. 

\begin{figure*}[h!]
   \centering
\includegraphics[width=\linewidth]{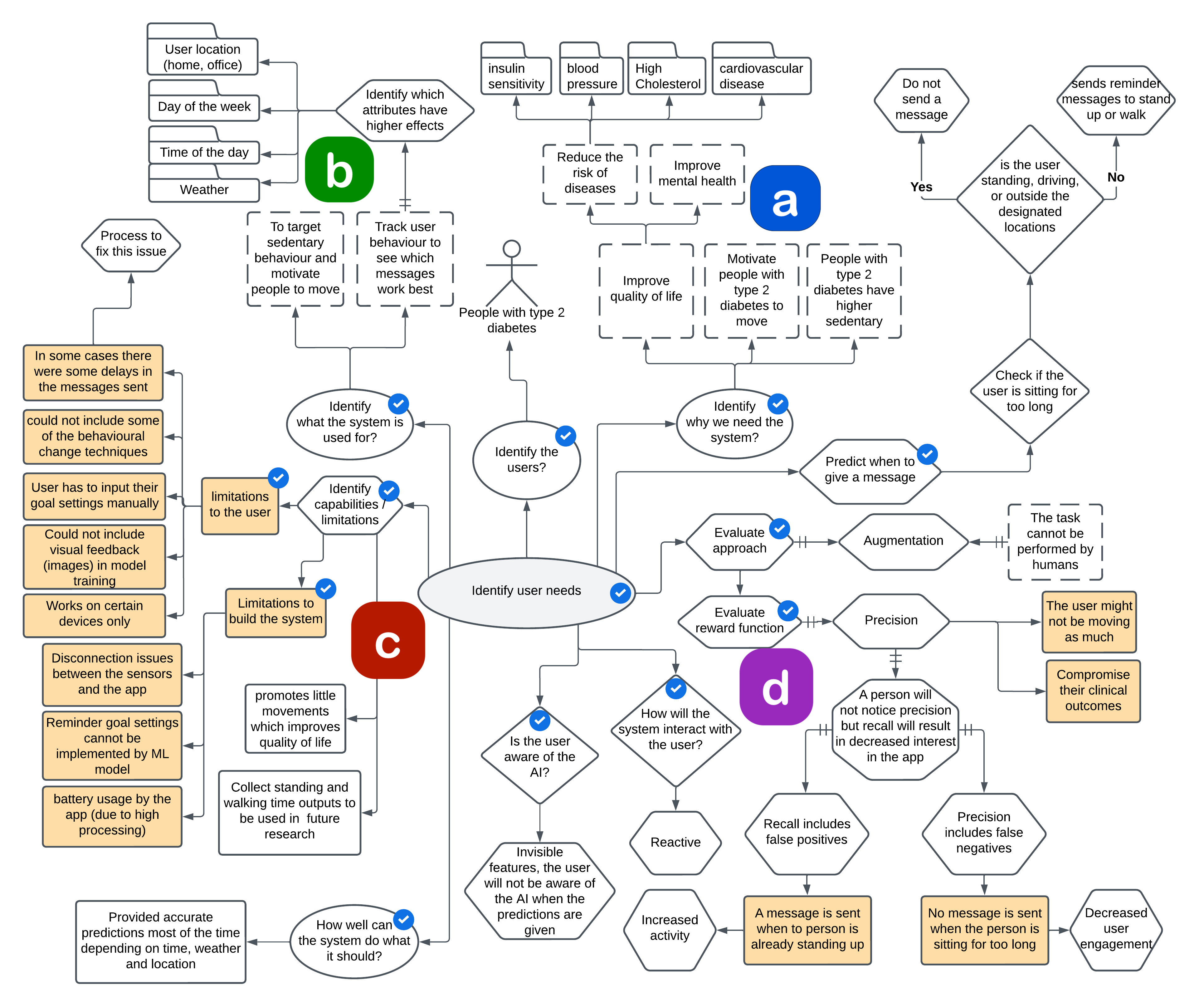}
\caption{User Needs for the \iMove~project}
\label{fig:UserNeeds-iMove}
\end{figure*}

When evaluating the need, it was evident that minimizing prolonged sitting times could reduce the risk of increased blood pressure, high cholesterol, and cardiovascular diseases, specifically for people with type 2 diabetes~\cite{daryabeygi2022development}. The idea behind using a mobile health application was to generate notifications by the system as to when to stand up. Therefore, the system will have to accurately predict if the user has been sitting too long and notify them to stand up or walk around. Since the application had to provide predictions, using AI (i.e., machine learning or deep learning) was a feasible solution. Figure~\ref{fig:UserNeeds-iMove} shows the system's different needs (part a), including improving mental health as an outcome of breaking up prolonged sitting times for people with T2D. Part \textbf{b} (the green colored box) of Figure~\ref{fig:UserNeeds-iMove} shows what the system is used for.


In part \textbf{c} (the red colored box) of Figure~\ref{fig:UserNeeds-iMove} we show the limitations and capabilities of the system.  There were several \emph{system limitations} that were identified early, and some of these limitations appeared as the project advanced in the later stages. Some of the limitations were easily addressed, e.g., delays in users receiving messages from the application and connection issues between the sensors and the application. Other system limitations were difficult to address. These included goal settings, as the model could not achieve this feature. Therefore, the user had to input the settings manually through the user interface. Also, part of the initial plan was to include behavioral change techniques (BCT), such as mood. However, this was later discarded as they found that including BCT features would increase the number of messages sent to the user, making it more of a hassle and an inconvenience to the end user. 

\begin{figure*}[h]
   \centering
\includegraphics[width=\linewidth]{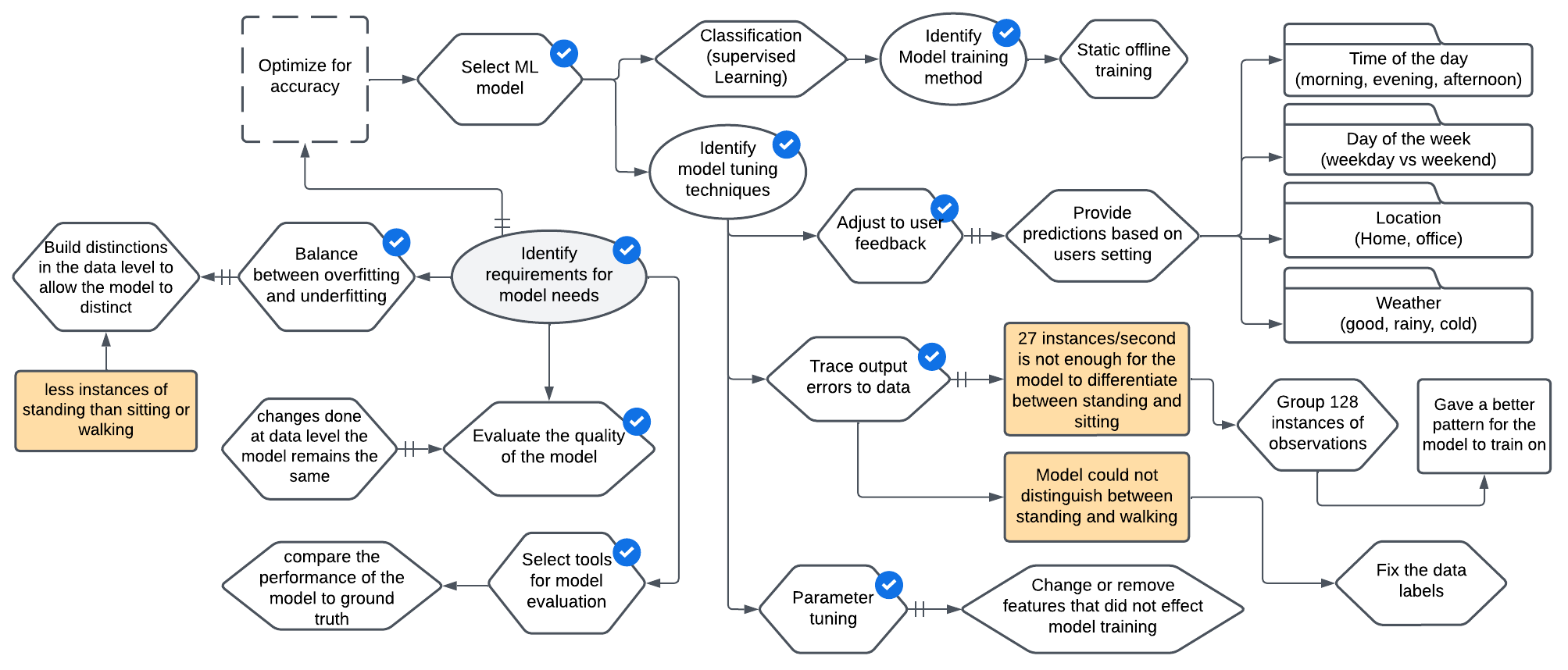}
\caption{Model Needs for the \iMove~project}
\label{fig:ModelNeeds-iMove}
\end{figure*}


When evaluating the \emph{reward function} or evaluation metric, we first set out to list the possible outcomes of getting either a False Positive (FP) or a False Negative (FN), as reflected in Part \textbf{d} (the purple colored box) of Figure~\ref{fig:UserNeeds-iMove}. We discussed these options and \emph{weighed out the trade-off}. An FP meant the system would send a message to remind the user to stand up or walk when they were already standing or walking. An FN happened when the person was sitting for too long, and the application would miss sending a reminder. In the case of an FP, the end user would lose interest in the system if incorrect notifications were sent and most likely become less interested in using the application altogether. Contrarily, when an FN occurs, the end user would not notice, as sitting is part of human nature. The downside to an FN was that the user would move less than required, which might complicate or delay their clinical goals. However, the project manager explained that from the user's perspective, having an FP had higher consequences than an FN, as an FN would not irritate the end user as much as having an FP. Thus, the aim was to minimize FP as much as possible when selecting the reward function.


\vspace{-0.7em}
\subsection{\iMove: Model Needs}~\label{sec:iMoveModelNeeds}
\vspace{-0.7cm}

The deep learning (DL) model was designed to \emph{optimize for accuracy}. Although we established in user needs that precision would be the evaluation metric, the predictions were evaluated based on accuracy only by the experts when building the model.
The selected \emph{DL classification model} was built to distinguish a person's movement behavior. Initially, the training was done offline by collecting data from sensors manually and using this data to train the model. Therefore, the model was not designed to learn and adjust to users' behavior while using the system. Figure \ref{fig:ModelNeeds-iMove} shows part of the model-needs model developed for the \iMove~project.

Most of the \emph{model tuning} happens at the data level for this project. When training the model, there were enough data instances, and the data had ground truth, thus making it easy to train the model. However, the experts expressed that while using the training data, the model could not distinguish between standing and walking instances. This issue was mainly due to the data collection method, as each data-collection instance had 27 sensor data records per second. One second's worth of data for each instance was not enough for the model to detect the users' movement when it came to standing and walking. To fix this issue, they had to increase the window to 128 records per second. This increase helped the model differentiate between standing and~walking.

\begin{figure*}[h!]
   \centering
\includegraphics[width=\linewidth]{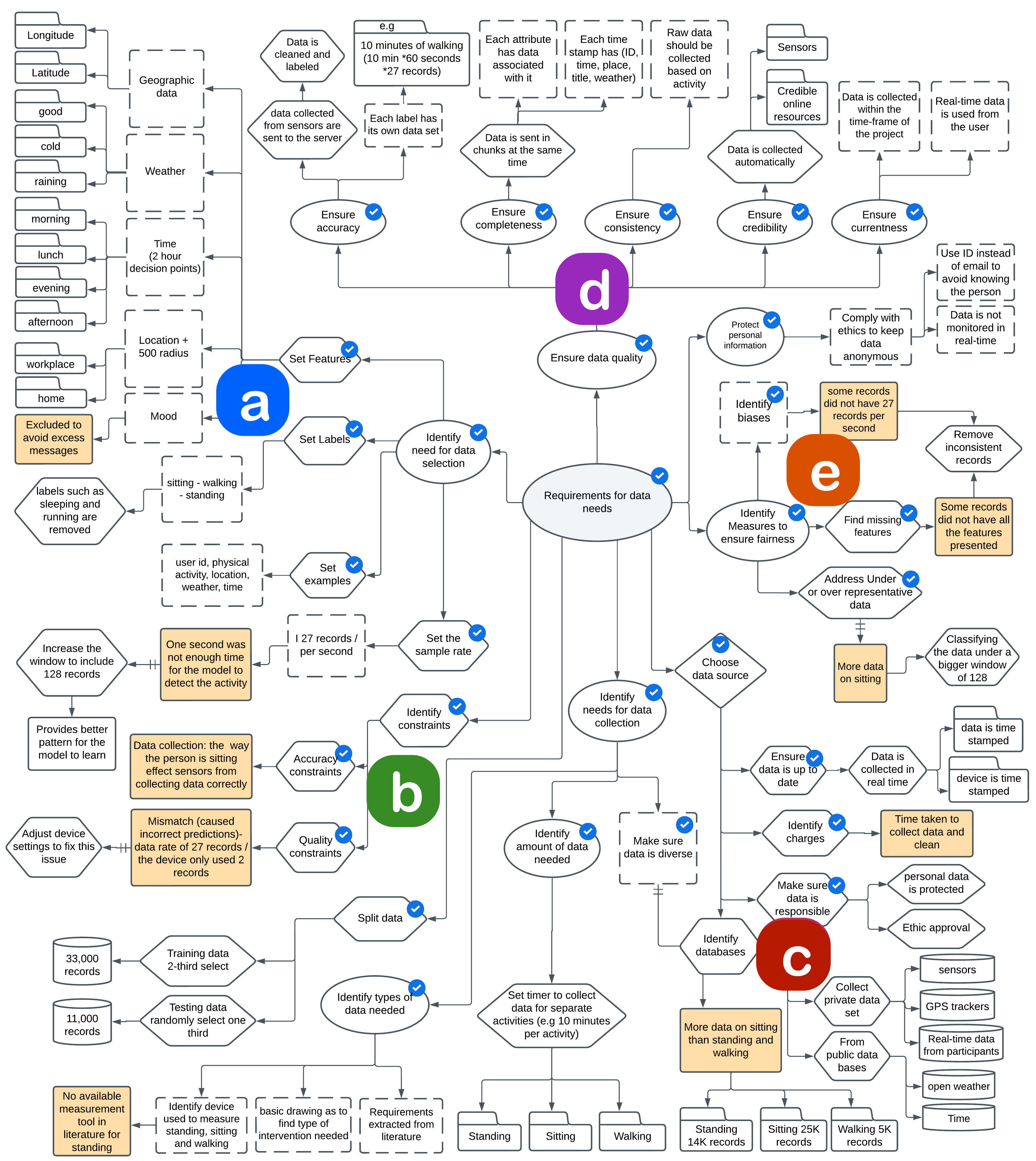}
\caption{Data Needs for the \iMove~project}
\label{fig:DataNeeds-iMove}
\end{figure*}

\vspace{-0.2cm}
\subsection{\iMove: Data Needs}~\label{sec:iMoveDataNeeds}
\vspace{-0.7cm}

The first requirement for data collection was to identify which features were needed, as shown in part \textbf{a} (blue coloured box) of Figure~\ref{fig:DataNeeds-iMove}. Features for the \iMove~included geographic data (longitude and latitude), location, weather, time of the day, and day of the week. The collected data was then labeled as sitting, walking, and standing. The data was cleaned, and some instances of running and sleeping were removed. An example of a row included the user ID, physical activity, location, sitting, standing, weather, and time. The initial sampling rate was 27 records per second, which was later updated to 128 records due to model limitations.

A private database was primarily built as the main \emph{data source}, yet some data was obtained from a public database, such as weather (open weather), as shown in part \textbf{c} (the red coloured box) of Figure~\ref{fig:DataNeeds-iMove}. For the private database, the data was collected in real-time in batches. Data was collected using Internet of Things (IoT) devices and sensors. Each sample was time-stamped by the device used to ensure the data was always up to date. Other resources, such as GPS trackers and time applications, were used. When collecting the data from sensors, they had to account for constraints, as shown in part \textbf{b} (the green coloured box) of Figure~\ref{fig:DataNeeds-iMove}. The first visible constraint was identified when some attributes were missing from the collected data. It was found that the way the user was sitting affected the sensors from collecting the data correctly. For example, data collected from a person sitting with one knee crossed over the other provided different measurements than someone sitting upright on a chair with both knees bent and feet on the floor. The other potential constraint was a mismatch between the data rate sent from sensors and the model. This mismatch resulted in predictions that were not accurate due to a fault in the system. The model was trained on 27 records per second, and the device collecting the data only transmitted two records per second. Modifications had to be made to the device to change the data rate and fix the fault in the system.
 
Requirements for \emph{data quality} included that the data should be accurate, complete, consistent, credible, and current. As displayed in part \textbf{d} (the purple coloured box) of Figure~\ref{fig:DataNeeds-iMove}. The data was collected from sensors and sent to the server as raw data. Thus, to ensure that the data was accurate, it had to be cleaned and labeled accordingly. Each label had its own dataset; i.e., collecting data on walking was done by setting a timer for 10 minutes (to collect 10*60*27 records) and labeling all the collected data as walking. Furthermore, data was collected based on the activity and sent in chunks, which helped build a more consistent dataset. For data completeness, the team experienced some data loss due to connectivity issues between the sensors and the device, so they had to ensure that the sent data matched the time stamps. For credibility, data were collected automatically from sensors, which they assumed were truthful. And last, data was seen as current since it was collected for the project for a period of time. 

The final step was to investigate whether the collected data was \emph{fair and inclusive}.  This is modeled in part \textbf{e} (the orange coloured box) of Figure~\ref{fig:DataNeeds-iMove}. First, the team had to apply for ethics and comply with the ethics application while collecting and using personal information. The team mentioned that it was challenging to make sure that the users' personal information was protected. They ended up providing each user with an ID and did not collect data from users in real-time to keep them anonymous. The next step was to find any missing features. When cleaning the data, they found that some instances did not have 27 records per second, which caused inconsistencies in the data. Also, the data on walking and standing was underrepresented, which caused issues in the accuracy of the predicted results. 

\vspace{-1.5em}

\section{\uppercase{Interviews}}~\label{sec:discussion}
\vspace{-2em}

We asked the \iMove~team for feedback on our proposed \REFramework~framework and its supporting collaborative tool. Below, we present their feedback on the framework and the tool and present some insights from applying our framework to the~\iMove~project.

\vspace{-0.5em}
\subsection{\REFramework~Framework Feedback}
\vspace{-1em}
\sectopic{Collaborative work.} Both experts highly rated the ability to work collaboratively on the platform.  The project manager felt that it was an excellent choice to use Confluence. They explain that \textit{``using confluence compared to other platforms that I had used before such as Jira was easier to learn''.} They also felt that having such a platform could help shorten the duration of the study and make everything much clearer.  

\sectopic{Visual Aspects.} Both experts said that the model's key advantages include providing a practical visual representation of the system's AI-based component requirements. Using yellow for limitations was favored by our participants. One participant mentioned that \textit{``limitations are negative, and having it colored yellow made it easier to distinguish''}. Also, using notations such as icons, database notations, and decisions was convenient as they could quickly point them out without needing to go back and check the legend. 

\sectopic{Limitations.} We asked our participants about the limitations of the catalog and modeling language. The project manager explained that they had issues understanding some of the terminologies used in the catalog. \textit{``I had some questions, and some aspects I did not understand as I am not from an engineering background, but you clarified those for me.''}. They suggested we provide examples to the catalog to explain the concepts and include a comment section to provide feedback when required. The major limitation of the model was that it could get slightly complex at times. For example, in the data needs model, a lot of information was presented, and it got difficult to keep track of the information provided. Suggestions to reduce the complexity of the final models included slicing the model into smaller parts or highlighting aspects of the diagram based on how it is categorized.
\subsection{Reflections}
\vspace{-1em}
\sectopic{Missing Requirements.} We noticed that the human-centred aspects presented in our catalog were not considered when originally building the~\iMove,  especially the user needs area.  Our participants said they had to re-consider most of the user needs only after viewing the models. Requirements such as evaluating the reward function were not discussed among the team members when building the project. For example, they did not consider the consequences of having a FP vs a FN in the app before starting the project. This reflects the need for a platform like ours to help guide the practitioners in eliciting and modeling requirements when building AI software.

\sectopic{New Roles, Collaboration and Communication.} AI paradigm has introduced new roles and requirements, which was not the case in traditional software engineering projects. For instance, we noticed that some of the requirements extracted from the project manager were not in line with the requirements taken from both the data scientist and software engineer. We used our framework and tool to assist the team in clarifying these crucial requirements issues. 

For example, when we collected requirements from the project manager, we established that precision should be used when evaluating the reward function. However, the data scientist working on the DL model focused on accuracy. Accuracy might not have been the best metric to use as high accuracy could have been achieved if the negative class was the dominant one~\cite{juba2019precision}. This would have caused issues in the results, as there was a visible class imbalance in the dataset. Class imbalance is when data from the major class is more represented than the other classes, which can cause the model to present biased results towards the dominant class~\cite{johnson2019survey,japkowicz2002class}. In this case, data on sitting was more represented than standing and walking, which might have contributed to the high accuracy. 

We observed that in some instances, the different team members were unaware of some aspects the other members were working on prior to introducing our tool. For example, the data scientist had to adjust the amount of data fed into the model to 128 records due to model limitations. The project manager became aware of this only after observing this limitation in our conceptual model. Also, the software engineer was only aware of setting up the sensors and collecting the data from the devices and was not able to provide much detail on what features were needed for the data, such as time and weather. 

\sectopic{Changing Requirements.} Another observation we made was that some of the \iMove~project requirements changed over time. For example, in terms of user needs, some of the system needs were found to be difficult to implement due to model limitations. The original plan was to include images when displaying the feedback to the end user. Over time, it became apparent that the model had limitations and could only provide textual messages. Thus, providing visual feedback became a limitation to the system rather than a need. Also, when collecting data requirements at the start of the project, it was not very clear how the data should be used to train the DL model. Initially, there was a lot of raw data to work with. However, the format and the diversity of the collected data were not adequate to use for training or testing the model. The data had to be modified throughout the project timeline on many occasions for it to become fit to train the selected model. This reflects the volatility and the exploratory nature of AI projects. 
 Thus a platform like ours facilitates discussions in the early project stages to establish foundational requirements, which all roles agree on. This would help deal with the changing requirements as well, and to see the impact of the changes on different parts of user needs, model needs or data needs.



\vspace{-2em}

\section{\uppercase{Conclusions}}~\label{sec:conclusion}

\vspace{-2em}
In this paper, we reported on a case study related to our framework and the related tool for eliciting and modeling requirements for human-centred AI. The tool is used to support our proposed framework based on the industrial human-centred guidelines, an analysis of studies obtained from the literature, and a user survey that we have conducted. The case study was conducted with three experts from a team working on a mobile health application targeted at type-2 diabetes patients. We found that most human-centred aspects were missing from the initial planning of the AI software solution. Also, we found that requirements change dramatically over time with AI software, and a more iterative approach to RE needs to be considered. In the future, we plan to conduct more case studies of our framework and the tool. \textcolor{black}{Specifically, we aim to select projects using different AI techniques, from different application domains, and at different stages of development. We want to investigate the difference in the RE process with and without our framework, in different settings. }


\vspace{-2em}


\bibliographystyle{apalike}
{\small
\bibliography{mybib_anon.bib}}

\begin{thebibliography}{}

\bibitem[Agarwal and Goel, 2014]{agarwal2014expert}
Agarwal, M. and Goel, S. (2014).
\newblock Expert system and it's requirement engineering process.
\newblock In {\em International Conference on Recent Advances and Innovations
  in Engineering (ICRAIE-2014)}, pages 1--4. IEEE.

\bibitem[Ahmad et~al., 2022]{ahmad2022requirements}
Ahmad, K., Abdelrazek, M., Arora, C., Bano, M., and Grundy, J. (2022).
\newblock Requirements engineering for artificial intelligence systems: A
  systematic mapping study.
\newblock {\em arXiv preprint arXiv:2212.10693}.

\bibitem[Ahmad et~al., 2023]{ahmad2023requirements}
Ahmad, K., Abdelrazek, M., Arora, C., Bano, M., and Grundy, J. (2023).
\newblock Requirements practices and gaps when engineering human-centered
  artificial intelligence systems.
\newblock {\em arXiv preprint arXiv:2301.10404}.

\bibitem[Ahmad et~al., 2021]{ahmad2021SLR}
Ahmad, K., Bano, M., Abdelrazek, M., Arora, C., and Grundy, J. (2021).
\newblock What’s up with requirements engineering for artificial intelligence
  systems?
\newblock In {\em 2021 IEEE 29th International Requirements Engineering
  Conference (RE)}, pages 1--12. IEEE.

\bibitem[Amaral et~al., 2020]{amaral2020ontology}
Amaral, G., Guizzardi, R., Guizzardi, G., and Mylopoulos, J. (2020).
\newblock Ontology-based modeling and analysis of trustworthiness requirements:
  Preliminary results.
\newblock In {\em International Conference on Conceptual Modeling}, pages
  342--352. Springer.

\bibitem[Amershi et~al., 2014]{amershi2014power}
Amershi, S., Cakmak, M., Knox, W.~B., and Kulesza, T. (2014).
\newblock Power to the people: The role of humans in interactive machine
  learning.
\newblock {\em {AI} Magazine}, 35(4).

\bibitem[Amyot and Mussbacher, 2011]{amyot2011user}
Amyot, D. and Mussbacher, G. (2011).
\newblock User requirements notation: the first ten years, the next ten years.
\newblock {\em JSW}, 6(5):747--768.

\bibitem[{Apple Developer}, 2020]{Apple2020}
{Apple Developer} (2020).
\newblock Human interface guidelines.
\newblock [online]
  https://developer.apple.com/design/human-interface-guidelines/machine-learning/overview/introduction/
  [Accessed 1 May 2020].

\bibitem[Belani et~al., 2019]{belani2019requirements}
Belani, H., Vukovic, M., and Car, {\v{Z}}. (2019).
\newblock Requirements engineering challenges in building {AI}-based complex
  systems.
\newblock In {\em 2019 IEEE 27th International Requirements Engineering
  Conference Workshops (REW)}, pages 252--255. IEEE.

\bibitem[Carrizo et~al., 2014]{carrizo2014systematizing}
Carrizo, D., Dieste, O., and Juristo, N. (2014).
\newblock Systematizing requirements elicitation technique selection.
\newblock {\em Information and Software Technology}, 56(6):644--669.

\bibitem[Chen et~al., 2022]{chen2022hint}
Chen, Q.~Z., Schnabel, T., Nushi, B., and Amershi, S. (2022).
\newblock Hint: Integration testing for ai-based features with humans in the
  loop.
\newblock In {\em 27th International Conference on Intelligent User
  Interfaces}, pages 549--565.

\bibitem[Dalpiaz et~al., 2016]{dalpiaz2016istar}
Dalpiaz, F., Franch, X., and Horkoff, J. (2016).
\newblock istar 2.0 language guide.
\newblock {\em arXiv preprint arXiv:1605.07767}.

\bibitem[Daryabeygi-Khotbehsara et~al., 2022]{daryabeygi2022development}
Daryabeygi-Khotbehsara, R., Shariful~Islam, S.~M., Dunstan, D.~W., Abdelrazek,
  M., Markides, B., Pham, T., and Maddison, R. (2022).
\newblock Development of an android mobile application for reducing sitting
  time and increasing walking time in people with type 2 diabetes.
\newblock {\em Electronics}, 11(19):3011.

\bibitem[Davey and Parker, 2015]{davey2015requirements}
Davey, B. and Parker, K.~R. (2015).
\newblock Requirements elicitation problems: a literature analysis.
\newblock {\em Issues in Informing Science and Information Technology},
  12:71--82.

\bibitem[Dempsey et~al., 2016]{dempsey2016sitting}
Dempsey, P.~C., Owen, N., Yates, T.~E., Kingwell, B.~A., and Dunstan, D.~W.
  (2016).
\newblock Sitting less and moving more: improved glycaemic control for type 2
  diabetes prevention and management.
\newblock {\em Current diabetes reports}, 16(11):1--15.

\bibitem[Ezzini et~al., 2023]{ezzini23ai}
Ezzini, S., Abualhaija, S., Arora, C., and Sabetzadeh, M. (2023).
\newblock Ai-based question answering assistance for analyzing natural-language
  requirements.
\newblock In {\em In Proceedings of the 45th International Conference on
  Software Engineering (ICSE'23)}.

\bibitem[Fazzini et~al., 2022]{Fazzini:22}
Fazzini, M., Khalajzadeh, H., Haggag, O., Li, Z., Obie, H., Arora, C., Hussain,
  W., and Grundy, J. (2022).
\newblock Characterizing human aspects in reviews of covid-19 apps.
\newblock In {\em 2022 IEEE/ACM 9th International Conference on Mobile Software
  Engineering and Systems (MobileSoft)}, pages 38--49.

\bibitem[Feldt et~al., 2018]{feldt2018ways}
Feldt, R., de~Oliveira~Neto, F.~G., and Torkar, R. (2018).
\newblock Ways of applying artificial intelligence in software engineering.
\newblock In {\em Proceedings of the 6th International Workshop on Realizing
  Artificial Intelligence Synergies in Software Engineering}, pages 35--41.

\bibitem[Fuentes-Fern{\'a}ndez et~al., 2010]{fuentes2010understanding}
Fuentes-Fern{\'a}ndez, R., G{\'o}mez-Sanz, J.~J., and Pav{\'o}n, J. (2010).
\newblock Understanding the human context in requirements elicitation.
\newblock {\em Requirements engineering}, 15(3):267--283.

\bibitem[Gervasi et~al., 2013]{gervasi2013unpacking}
Gervasi, V., Gacitua, R., Rouncefield, M., Sawyer, P., Kof, L., Ma, L., Piwek,
  P., Roeck, A.~d., Willis, A., Yang, H., et~al. (2013).
\newblock Unpacking tacit knowledge for requirements engineering.
\newblock In {\em Managing requirements knowledge}, pages 23--47. Springer.

\bibitem[Gon{\c{c}}alves et~al., 2019]{gonccalves2019understanding}
Gon{\c{c}}alves, E., de~Oliveira, M.~A., Monteiro, I., Castro, J., and
  Ara{\'u}jo, J. (2019).
\newblock Understanding what is important in istar extension proposals: the
  viewpoint of researchers.
\newblock {\em Requirements Engineering}, 24(1):55--84.

\bibitem[{Google Research}, 2019]{GooglePair2019}
{Google Research} (2019).
\newblock The people + \uppercase{AI} guidebook.
\newblock [online] Available at: https://research.google/teams/brain/pair/
  [Accessed 1 April 2020].

\bibitem[Gruber et~al., 2017]{gruber2017integrated}
Gruber, K., Huemer, J., Zimmermann, A., and Maschotta, R. (2017).
\newblock Integrated description of functional and non-functional requirements
  for automotive systems design using sysml.
\newblock In {\em 2017 7th IEEE International Conference on System Engineering
  and Technology (ICSET)}, pages 27--31. IEEE.

\bibitem[Grundy, 2021]{grundy2021impact}
Grundy, J.~C. (2021).
\newblock Impact of end user human aspects on software engineering.
\newblock In {\em ENASE}, pages 9--20.

\bibitem[Holmquist, 2017]{holmquist2017intelligence}
Holmquist, L.~E. (2017).
\newblock Intelligence on tap: artificial intelligence as a new design
  material.
\newblock {\em interactions}, 24(4):28--33.

\bibitem[Hong et~al., 2021]{hong2021planning}
Hong, M.~K., Fourney, A., DeBellis, D., and Amershi, S. (2021).
\newblock Planning for natural language failures with the ai playbook.
\newblock In {\em Proceedings of the CHI Conference on Human Factors in
  Computing Systems}, pages 1--11.

\bibitem[Japkowicz and Stephen, 2002]{japkowicz2002class}
Japkowicz, N. and Stephen, S. (2002).
\newblock The class imbalance problem: A systematic study.
\newblock {\em Intelligent data analysis}, 6(5):429--449.

\bibitem[Johnson and Khoshgoftaar, 2019]{johnson2019survey}
Johnson, J.~M. and Khoshgoftaar, T.~M. (2019).
\newblock Survey on deep learning with class imbalance.
\newblock {\em Journal of Big Data}, 6(1):1--54.

\bibitem[Juba and Le, 2019]{juba2019precision}
Juba, B. and Le, H.~S. (2019).
\newblock Precision-recall versus accuracy and the role of large data sets.
\newblock In {\em Proceedings of the AAAI conference on artificial
  intelligence}, volume~33, pages 4039--4048.

\bibitem[Kennerly and Kirk, 2018]{kennerly2018physical}
Kennerly, A.-M. and Kirk, A. (2018).
\newblock Physical activity and sedentary behaviour of adults with type 2
  diabetes: a systematic review.
\newblock {\em Practical Diabetes}, 35(3):86--89g.

\bibitem[Khomh et~al., 2018]{khomh2018software}
Khomh, F., Adams, B., Cheng, J., Fokaefs, M., and Antoniol, G. (2018).
\newblock Software engineering for machine-learning applications: The road
  ahead.
\newblock {\em IEEE Software}, 35(5):81--84.

\bibitem[Lapouchnian, 2005]{lapouchnian2005goal}
Lapouchnian, A. (2005).
\newblock Goal-oriented requirements engineering: An overview of the current
  research.
\newblock {\em University of Toronto}, 32.

\bibitem[{Louis Dorard}, ]{MLCanvas}
{Louis Dorard}.
\newblock The machine learning canvas.
\newblock https://www.louisdorard.com/machine-learning-canvas. Accessed [March,
  2020].

\bibitem[Lu et~al., 2022]{lu2022software}
Lu, Q., Zhu, L., Xu, X., Whittle, J., Douglas, D., and Sanderson, C. (2022).
\newblock Software engineering for responsible ai: An empirical study and
  operationalised patterns.
\newblock In {\em 2022 IEEE/ACM 44th International Conference on Software
  Engineering: Software Engineering in Practice (ICSE-SEIP)}, pages 241--242.
  IEEE.

\bibitem[Maguire, 2001]{maguire2001methods}
Maguire, M. (2001).
\newblock Methods to support human-centred design.
\newblock {\em International journal of human-computer studies},
  55(4):587--634.

\bibitem[Mart{\'\i}nez-Fern{\'a}ndez et~al., 2022]{martinez2022software}
Mart{\'\i}nez-Fern{\'a}ndez, S., Bogner, J., Franch, X., Oriol, M., Siebert,
  J., Trendowicz, A., Vollmer, A.~M., and Wagner, S. (2022).
\newblock Software engineering for ai-based systems: a survey.
\newblock {\em ACM Transactions on Software Engineering and Methodology
  (TOSEM)}, 31(2):1--59.

\bibitem[{Microsoft}, 2022]{Microsoft2022}
{Microsoft} (2022).
\newblock Guidelines for human-ai interaction.
\newblock [online]
  https://www.microsoft.com/en-us/research/project/guidelines-for-human-ai-interaction/
  [Accessed 1 Feb 2022].

\bibitem[Moody, 2009]{moody2009physics}
Moody, D. (2009).
\newblock The “physics” of notations: toward a scientific basis for
  constructing visual notations in software engineering.
\newblock {\em IEEE Transactions on software engineering}, 35(6):756--779.

\bibitem[Nalchigar et~al., 2021]{nalchigar2021modeling}
Nalchigar, S., Yu, E., and Keshavjee, K. (2021).
\newblock Modeling machine learning requirements from three perspectives: a
  case report from the healthcare domain.
\newblock {\em Requirements Engineering}, 26(2):237--254.

\bibitem[Nuseibeh and Easterbrook, 2000]{nuseibeh2000requirements}
Nuseibeh, B. and Easterbrook, S. (2000).
\newblock Requirements engineering: a roadmap.
\newblock In {\em Proceedings of the Future of Software Engineering
  Conference}, pages 35--46.

\bibitem[Ries et~al., 2021]{ries2021mde}
Ries, B., Guelfi, N., and Jahic, B. (2021).
\newblock An {MDE} method for improving deep learning dataset requirements
  engineering using alloy and {UML}.
\newblock In {\em Proceedings of the 9th International Conference on
  Model-Driven Engineering and Software Development}, pages 41--52.

\bibitem[Schmidt, 2020]{schmidt2020interactive}
Schmidt, A. (2020).
\newblock Interactive human centered artificial intelligence: a definition and
  research challenges.
\newblock In {\em Proceedings of the International Conference on Advanced
  Visual Interfaces}, pages 1--4.

\bibitem[Sculley et~al., 2015]{sculley2015hidden}
Sculley, D., Holt, G., Golovin, D., Davydov, E., Phillips, T., Ebner, D.,
  Chaudhary, V., Young, M., Crespo, J.-F., and Dennison, D. (2015).
\newblock Hidden technical debt in machine learning systems.
\newblock In {\em Advances in neural information processing systems}, pages
  2503--2511.

\bibitem[Shneiderman, 2022]{shneiderman2022human}
Shneiderman, B. (2022).
\newblock {\em Human-Centered AI}.
\newblock Oxford University Press.

\bibitem[Stephenson et~al., 2017]{stephenson2017using}
Stephenson, A., McDonough, S.~M., Murphy, M.~H., Nugent, C.~D., and Mair, J.~L.
  (2017).
\newblock Using computer, mobile and wearable technology enhanced interventions
  to reduce sedentary behaviour: a systematic review and meta-analysis.
\newblock {\em International Journal of Behavioral Nutrition and Physical
  Activity}, 14(1):1--17.

\bibitem[Sutcliffe and Sawyer, 2013]{sutcliffe2013requirements}
Sutcliffe, A. and Sawyer, P. (2013).
\newblock Requirements elicitation: Towards the unknown unknowns.
\newblock In {\em 2013 21st IEEE International Requirements Engineering
  Conference (RE)}, pages 92--104. IEEE.

\bibitem[Villamizar et~al., 2021]{villamizar2021requirements}
Villamizar, H., Escovedo, T., and Kalinowski, M. (2021).
\newblock Requirements engineering for machine learning: A systematic mapping
  study.
\newblock In {\em 2021 47th Euromicro Conference on Software Engineering and
  Advanced Applications (SEAA)}, pages 29--36. IEEE.

\bibitem[Zowghi and Coulin, 2005]{zowghi2005requirements}
Zowghi, D. and Coulin, C. (2005).
\newblock Requirements elicitation: A survey of techniques, approaches, and
  tools.
\newblock In {\em Engineering and managing software requirements}, pages
  19--46. Springer.

\end{thebibliography}






\end{document}